\documentclass[11pt]{article}
\usepackage{amsmath,amsthm,latexsym,amssymb,amsfonts,epsfig}

\newtheorem{lemm}{Lemma}[section]

\newtheorem{defi}[lemm]{Definition}

\newcommand{\R}{\mathbb{R}}                  
\newcommand{\C}{\mathbb{C}}                  
\newcommand{\N}{\mathbb{N}}                  

\DeclareMathOperator{\im}{Im}                          

\newcommand{\kwa}{\mathfrak{A}}           

\newcommand{\hilb}{\mathcal{H}}               
\newcommand{\one}{\text{\bf 1}}               

\newcommand{\U}{{\cal U}}

\newcommand{\Cyl}{{\rm Cyl}}
\newcommand{\ka}{{\bf k}}

\newcommand{\gw}[1]{{#1}^\ast}

\begin{document}
\title{Background independent quantizations: the scalar field I}
\author{Wojciech Kami\'nski$^1$,
        Jerzy Lewandowski$^{1,2}$\thanks{lewand@fuw.edu.pl}
and Marcin Bobie{\'n}ski$^3$}
\date{\it 1. Instytut Fizyki Teoretycznej,
Uniwersytet Warszawski, ul. Ho\.{z}a 69, 00-681 Warszawa, Poland\\
2. Perimeter Institute for Theoretical Physics, 31 Caroline Street North,
Waterloo, Ontario N2L 2Y5, Canada\\
3. Wydzia{\l} Matematyki, Informatyki i Mechaniki
 Uniwersytetu Warszawskiego ul. Banacha 2, 02-097 Warszawa, Poland} 
\maketitle
\begin{abstract}
We are concerned with the issue of quantization
of a scalar field in a diffeomorphism invariant  manner. We apply the method
used  in Loop Quantum Gravity.
It relies on a specific  choice of scalar field variables referred to as
the polymer variables. The
quantization, in our formulation, amounts to introducing the `quantum'
{\it polymer *-star algebra} and looking for  positive
linear functionals, called  states. Assumed in our paper homeomorphism
invariance allows to determine a complete class of the states.
Except one, all of them are new. In this letter we outline
the main steps and conclusions, and present
the results:  the GNS representations, characterization of those states
which lead to essentially self adjoint momentum operators (unbounded),
identification of the equivalence classes of the representations as well
as of the  irreducible ones. The algebra and topology of the problem,
the derivation and all the technical details are contained in the
paper-part II.
\end{abstract}
\section{Motivation}\label{Motivation}
The phrase ``background independent theory'' means in Physics a
theory defined on a bare manifold endowed with no extra structure
like geometry or fixed coordinates. A prominent example is the
theory of matter fields coupled to Einstein's gravity. In the case
of a background independent classical theory it is natural to
assume the background independence in a corresponding quantum
theory.
 A profound polemic devoted to that issue
can be found in recent paper of Smolin \cite{l1}.\footnote{As far
as we are concerned, it is conceivable, that a background
independent quantum theory can be derived even from a background
dependent framework. Another possibility is, that  the classical
limit of the theory (the classical GR for example) has more
symmetries (the diffeomorphisms) then the underlying  quantum
theory. Therefore, we are not in the position to claim that every
background dependent approach to a quantization of a background
independent theory is wrong. Nonetheless, the first thing one
should do is to try without introducing  extra structures.} The
canonical formulation of the field theory relies on the $3+1$
decomposition of space-time into the `space' $M$ and `time' $\R$.
Then, the background independence implies invariance with respect
to the diffeomorphisms of $M$. The invariance concerns in
particular any  matter fields in question: they have to be
quantized in an often new, background independent way.

In this letter and the accompanying paper \cite{kl}  we are
concerned with the issue of a diffeomorphism invariant quantization
of a scalar field.
We follow the approach introduced by Thiemann \cite{t1}.  It was
motivated by and suited to be compatible with Loop Quantum Gravity \cite{lqg}
(LQG itself is not in the scope of this paper). Later, that approach was
called the {\it polymer
representation}  in \cite{als}.\footnote{The reason for that name, is that
the polymer excitations  of the quantum scalar field
serve as a toy model of the genuinely polymer,  supported on graphs
excitations of quantum geometry \cite{al2,lqg}.} It consists in a specific
choice of scalar field variables --the polymer variables--
as the starting point for a quantization. The quantization, in
our formulation, amounts to introducing the `quantum' {\it polymer *-star
algebra} and looking for its representations. We focus here on the so called
GNS representations defined by a positive linear functional, called a state.
An assumption of the homeomorphism invariance actually allows to
derive a class of the states. In this letter we outline
the main steps and conclusions of the derivation and present
the results:  the unique class of  states on the polymer *-algebra
which satisfy our assumptions. We also present the representations
they define and  characterize those states which lead to essentially
self adjoint momentum operators (unbounded). We identify the equivalence
classes of the representations as well as  irreducible ones.
The technical and detailed derivation of the results is contained
in \cite{kl}. The problem we have addressed and the methods we have used
were inspired by the previous  study \cite{lost0,lost}  of representations
of the Sahlmann quantum holonomy-flux algebra in LQG.

\section{Classical polymer Lie algebra $\kwa_{\rm cl}$}\label{classical}
The classical scalar field in the canonical approach consists of a
pair of fields $(\phi, \pi)$ defined on a $N$-real dimensional
manifold $M$, where: $\phi\in C^\infty(M,\R)$ where
$C^\infty(M,\R)$ stands for the space of the smooth real valued
functions defined on $M$, whereas $\pi$, called canonical
momentum, is a scalar density of the weight $1$. The momentum
$\pi$ can be expressed by  a function $\tilde{\pi}:U\rightarrow
\R$ defined on an arbitrary region $U\subset M$ equipped with
coordinates $(x^1,...,x^N)$. The function, however, depends on the
coordinates in the following way:  if $(x'^1,...,x'^N)$ is another
coordinate system defined on $U$, then the corresponding momentum
function $\tilde{\pi}'$ is such that at every $x\in U$
\begin{equation}
\tilde{\pi}(x)d^Nx\ =\ \tilde{\pi}'(x)d^Nx'.
\end{equation}
The fields $\phi$ and $\pi$ are called canonically conjugate in the sense
od the Poisson bracket, that is usually wrote as
\begin{equation}\label{poisson}
\{\phi(x), \tilde{\pi}(y)\}\ =\ \delta_x(y).
\end{equation}
That Poisson bracket will be encoded in the Lie bracket of a Lie
algebra.

A choice of  those functions of the fields $(\phi,\pi)$ which next
will be  turned into quantum operators directly, is crucial for
the quantization procedure. They will be {\it polymer variables}
defined below.

A {\it  polymer position variable}  $h_{k,x}$ is labelled by a
point $x\in M$ and a number $k$, as the evaluation function
(scaled and exponentiated),
\begin{equation}\label{h}
h_{k,x}(\phi,\pi)\ =\  e^{ik\phi(x)}.
\end{equation}
A {\it polymer momentum variable} $\pi_U$ is labelled by a region $U\subset M$,
\begin{equation}\label{piu}
\pi_U\ =\ \int_U\pi,
\end{equation}
where the integral is well defined because $\pi$ has the properties
of a signed measure.
\medskip

The polymer position and, respectively, momentum variables span a
{\it polymer Lie algebra  $(\kwa_{\rm cl}, \{\cdot,\cdot\})$}\footnote{Each
symbol $`\cdot'$ in the bracket
means a spot for an arbitrary element of $\kwa$.}:
\begin{equation}
\kwa_{\rm cl}\ =\ \left\{\sum_{i=1}^n a_i h_{k_i,x_i} +
\overline{\sum_{i=1}^n a_i h_{k_i,x_i}} + \sum_{j=1}^mb_i\pi_{U_i}\ :\
a_i\in \C, b_j\in\R, U_j\in {\U}\right\}
\end{equation}
where ${\cal U}$ is some suitably fixed family of sets (to be specified later),
and the Lie bracket is the Poisson bracket
\begin{align}
\{a h_{k,x} + \bar{a} h_{-k,x}\,,\,\pi_U\}\ &=\ 
\left\{\begin{array}{ll} ik(ah_{k,x}-\bar{a}h_{-k,x}),
& {\rm if\ } x\in U\\
0,& {\rm if\ } x\notin U\\
\end{array}\right.\\
\{ah_{k,x}+\bar{a}h_{-k,x}\,,\,a'h_{k',x'}+\bar{a}'h_{-k',x'}\}\ =\ 
&0\ =\ \{\pi_U\,,\,\pi_{U'}\}
\end{align}

\section{Quantum polymer *-algebra $\kwa$}\label{quantum}
Now, we will  turn the polymer  position  and momentum variables
into {\it quantum} operators. Heuristically, we just
want to replace all the $h_{k,x}$ and $ \pi_U$ variables by quantum
operators   $\hat{h}_{k,x}$ and  $\hat{\pi}_U$.
We express  that well understood procedure by  a mathematically
complete definition.

First, consider the complexification $(\kwa_{\rm cl}^\C,
\{\cdot,\,\cdot\})$, of the polymer Lie algebra  $(\kwa_{\rm cl},
\{\cdot,\,\cdot\})$.

Next, consider the huge space,
\begin{equation}
e^{\otimes \kwa_{\rm cl}^\C}\ =\
\bigoplus_{n=0}^{\infty} (\kwa_{\rm cl}^\C)^{\otimes n}
\end{equation}
where
\begin{equation}
(\kwa_{\rm cl}^\C)^0:=\C.
\end{equation}
It has  the natural structure a complex, associative
(that is with an operation of multiplication) algebra
defined by the complex vector space structure and the operation
$\otimes$.
There is also naturally defined anti-isomorphism-convolution
$\gw{\cdot}$ in it, such that
\begin{equation}
\gw{(a_1\otimes\cdots\otimes a_n)}=\overline
{a_n}\otimes\cdots\otimes \overline{a_1},
\end{equation}
where  $\ \bar{{}}\ $  is the complex conjugation.

Next, consider the following subset of `relations'
\begin{equation}
\{\,a\otimes b- b\otimes a-i\{a,b\}\, :\, a,b\in \kwa_{\rm cl}^\C\,\},
\end{equation}
and  the corresponding double sided ideal\footnote{
That is the vector subspace of $e^{\otimes \kwa_{\rm cl}^\C}$
spanned by all the elements of the form $A\otimes(a\otimes b - b\otimes a -
i\{a,b\})\otimes B$ where $a,b\in \kwa^\C_{\rm cl}$
and $A,B\in e^{\otimes \kwa_{\rm cl}^\C}$ are all arbitrary.}
$J$ they generate.
Notice, that $J$ is preserved by the convolution $\gw{\cdot}$, hence
$\gw{\cdot}$ passes to the quotient $\bigoplus_{n=0}^{\infty}
(\kwa_{\rm cl}^\C)^{\otimes n}/J$ (and is denoted by the same symbol
$\gw{\cdot}$).

Finally, define:

\begin{defi}
The polymer *-algebra $(\kwa, \gw{\cdot})$ is the associative  *-algebra
\begin{equation}
\kwa\ =\ \bigoplus_{n=0}^{\infty} (\kwa_{\rm cl}^\C)^{\otimes n}/J.
\end{equation}
\label{kwaa}
\end{defi}

We will be using the following notation: an element
  $[a_1\otimes\dots\otimes a_n]\in \kwa$
corresponding to $a_1\otimes\dots\otimes a_n\in
e^{\otimes \kwa_{\rm cl}^\C}\ $, where
$a_1,\dots,a_n\in \kwa_{\rm cl}^\C$ will be denoted by
 $\hat{a}_1\dots \hat{a}_n$, in particular
\begin{align}
\hat{h}_{k,x}\ =\ [h_{k,x}]\in \kwa,\ \ \ \hat{\pi}_U \ =\ [\pi_U]\in \kwa\\
\hat{a}_1\dots \hat{a}_n\ =\ [a_1\otimes\dots\otimes a_n]\in \kwa.
\end{align}

Therefore indeed, the elements of the polymer *-algebra $\kwa$
can be thought of, as
polynomials in all the {\it quantum position polymer variables $\hat{h}_{k,x}$}
 and all the  {\it quantum momentum polymer variables  $\hat{\pi}_U$}.

The only non-vanishing  commutators between the generators of $\kwa$ are,
\begin{equation}\label{[h,pi]}
\hat{h}_{k,x}\hat{\pi}_U -\hat{\pi}_U\hat{h}_{k,x}\ =\
\left\{\begin{array}{ll} k\hat{h}_{k,x},&
{\rm if}\ \ x\in U\\
0,& {\rm if}\ \ x\notin U\\
\end{array}\right.\\
\end{equation}
for every $U\in {\cal U}$, $x\in M$ and $k\not= 0$.

Notice, that the zeroth order polynomials in the  quantum polymer position and
momentum variables
are also contained in $\kwa$. The are elements of the subspace
$\C\subset e^{\otimes\kwa^\C_{\rm cl}}$.  In particular, the element $1\in \C$,
defines an element $\hat{1}\in \kwa$, the unity of the polymer *-algebra
$\kwa$.

\section{States on $\kwa$: the nature of the problem}\label{formulation}
A complete quantization of the polymer variables amounts to
finding a representation of the polymer *-algebra $\kwa$ in the
algebra of operators defined in a Hilbert space. In this paper we
will be concerned with so called `GNS representations' defined by
so called  `states'  on $\kwa$. Recall, that a {\it state} on a
unital (i.e. containing an element
  $\hat{1}$ such that $\hat{1}A =A= A\hat{1}$ for every $A\in \kwa$)
$*$-algebra $\kwa$ is a linear functional
\begin{equation}
\kwa\ni A\ \mapsto\ \langle A\rangle\in \C,
\end{equation}
such that for every $A\in\kwa$
\begin{equation}\label{state1}
\langle\gw{A}A\rangle\ \ge 0, \ \ \ \  {\rm and}\ \ \ \
\langle\hat{1}\rangle\ =\ 1.
\end{equation}
Using the common jargon, the number $ \langle A\rangle$
is referred to as the `expectation value'.

The difficulty in defining a state on the polymer *-algebra $\kwa$
just by assigning a number to a general product of the generators
(the products  of the quantum momenta (positions) only
 are included in the case $n=0$ ($m=0$)),
\begin{equation}
\prod_{i=1}^n \hat{h}_{k_i,x_i}\prod_{j=1}^m\hat{\pi}_{U_j}\ \mapsto
\langle\prod_{i=1}^n \hat{h}_{k_i,x_i}\prod_{j=1}^m\hat{\pi}_{U_j}\rangle
\end{equation}
 and attempting to extend that
assignment by the linearity to the entire
$\kwa$  consists in the both:
\begin{itemize}
\item{} satisfying identities in the algebra -- all of them have to pass
to the expectation values,
\item{} ensuring the positivity.
\end{itemize}

Examples of the identities are:
\begin{equation}\label{identity}
\hat{\pi}_{U\cup U'}\ =\ \hat{\pi}_U + \hat{\pi}_{U'} - \hat{\pi}_{U\cap U'}
\end{equation}
which holds whenever $U,\, U',\,  U\cup U',\, U\cap U'\,\in {\cal U}$, as well
as the commutation relations (\ref{[h,pi]}) and obviously
the commutativity of the quantum polymer position (momentum)  variables among
themselves.

Nonetheless, there is a quite simple example of a state:
\medskip

\noindent{\bf Example} \cite{t1,als} Going back to the definition
(\ref{h}) of the polymer position variable, it is reasonable to
assume about an eventual representation of the polymer *-algebra
$\kwa$, that for  each point $x\in M$, the given family  of the
classical variables $e^{ik\phi(x)}$ correspond     to a
$1$-dimensional group of unitary operators. In terms of a state
$\langle\cdot\rangle$  on the polymer *-algebra $\kwa$, it is
equivalent to assuming that every product
$\hat{h}_{k,x}\hat{h}_{k',x}$  will be indistinguishable from
$\hat{h}_{k+k',x}$, that is
\begin{equation}
\langle A\cdot(\hat{h}_{k,x}\hat{h}_{k',x} - \hat{h}_{k+k',x})\cdot B\
 \rangle =0,\label{hh}\
\end{equation}
for every $A,B\in \kwa$, $x\in M$ and $k,k'\not=0$,  where, exclusively
in this equality, we also define $\hat{h}_{0,x}=\hat{1}$.

The following properties
\begin{align}
\langle A\cdot\pi_U\rangle\ &=\ 0  \ \ \ \ \ {\rm for \ \ \ every}\ \ U\in \U,
\label{example1} \\
\langle \prod_{i=1}^n \hat{h}_{k_i,x_i}\rangle\ &=\ 0
\ \ \ {\rm if} \ \ \ \left(x_i\not= x_j\ \  {\rm whenever}\ \  i\not= j\right)
\label{example2} .
\end{align}
together with the condition (\ref{hh}) define a unique state
$\langle\cdot\rangle$ on the polymer *-algebra $\kwa$. This state
is well known,  and used in LQG for the description of the scalar
field interacting with the quantum geometry.
\medskip

Thus far we have not specified the labelling family ${\cal U}$,
and Example above is not sensitive on a choice we make.

Henceforth, we will be  assuming that $\U$ consists of all the
subsets of $M$ which are simultaneously:  open, bounded and
triangulable (later in this section the last condition will be
strengthen).

The diffeomorphisms of $M$ act naturally on the classical fields $(\phi,\pi)$,
and that action passes to the polymer *-algebra $\kwa$.
 Given a diffeomorphism
$\varphi:M\rightarrow M$, the corresponding automorphism
$\varphi*:\kwa\rightarrow\kwa$ acts on the generators of $\kwa$ by
the action of $\varphi$ on the labelling elements and subsets of
$M$,
\begin{equation}\label{diff}
\varphi*\hat{h}_{k,x}\ =\ \hat{h}_{k,\varphi(x)},\ \ \ \
\varphi*\hat{\pi}_{U}\ =\ \hat{\pi}_{\varphi(U)}.
\end{equation}
Obviously, the state in Example is invariant with respect
to that action,
\begin{equation}
\langle\varphi*A\rangle\ =\ \langle A\rangle\, \ \ \ \
{\rm for\ \ every}\ \ A\in\kwa.
\end{equation}
The action (\ref{diff}) of the diffeomorphisms naturally extends
to the action of homeomorphisms of $M$ in $\kwa$. The state
(\ref{hh},\, \ref{example1},\,\ref{example2}) is also
homeomorphism invariant.

In the view of the symmetries of the state
(\ref{hh},\,\ref{example1},\,\ref{example2}) it
is natural to ask if there are other  homeomorphism invariant states
defined on the polymer *-algebra $\kwa$.
Derivation of such states was the main goal of the current and the
coming paper \cite{kl}. We discuss the results below whereas the
details and exact proofs will be published in the second paper.

\section{Homeomorphism invariant states on $\kwa$: properties}
\label{necessary}
A first surprising consequence of assuming the homeomorphism invariance
of a state $\langle\cdot \rangle$ on the polymer *-algebra $\kwa$,
is that for arbitrary quantum polymer  momentum
 variable $\hat{\pi}_U$ and every homeomorphism  $\varphi:M\rightarrow M$,
necessarily the following identity holds
\begin{equation}\label{ball}
\langle \gw{(\hat{\pi}_U -\hat{\pi}_{\varphi(U)})}
(\hat{\pi}_U -\hat{\pi}_{\varphi(U)}) \rangle\ =\ 0
\end{equation}
(the momenta are real, hence the $\gw{}$ does not change them, but
we put it to emphasis  the relation with (\ref{state1}).)

The careful analysis of the equality (\ref{ball}) allows us to determine for
every $n$-tuple of the quantum polymer momentum variables
$\hat{\pi}_{U_1},\ldots,\hat{\pi}_{U_k}\in
\kwa$, a possible  expectation value of the product up to a constant factor.
The result is
\begin{equation}\label{prodpi}
\langle \prod_{i=1}^n\hat{\pi}_{U_i}\rangle\ =
\ a_n \prod_{i=1}^n\chi_{\rm E}(U_i),
\end{equation}
where $\chi_{\rm E}(U_i)$ stands for the Euler characteristic of the set
$U_i$, and the constant $a_n$ depends only on $n$.

The sequence of constants $(a_n)_{n=1}^\infty$ in (\ref{prodpi})
can not be defined arbitrarily, as we attempt to construct a
state, because of the inequalities (\ref{state1}). A somewhat
formal, at a first sight, but quite convenient way of translating
(\ref{state1}) into a condition on the sequence
$(a_i)_{i=1}^\infty$, is by using polynomials defined on $\R$. Let
$C[t]$ be the space of all the polynomials (includding the
$0$th-order ones) with complex coefficients, defined on $\R$. It
has the natural *-algebra structure with $\gw{\cdot}$ being the
complex conjugation. The statement is, that  the infinite sequence
of the numbers $(a_i)_{i=1}^\infty$ defined by  (\ref{prodpi})
satisfies all the inequalities implied by the condition
(\ref{state1}) imposed on the subalgebra of the polymer *-algebra
$\kwa$ generated by all the quantum polymer momentum variables, if
and only if there is  a state $\mu$ on the *-algebra $C[t]$, such
that for every $n\in\N$,
\begin{equation}\label{a_n}
\mu(t^n)\ =\ a_n,
\end{equation}
where $t:\R\rightarrow \R$ is the identity function $t(t')=t'$.

This formulation of the conditions on the sequence  $(a_i)_{i=1}^\infty$
is particularly helpful  in constructing  solutions. In fact, each
probability measure $d\mu$ defined on $\R$ such that every polynomial
is integrable, defines a state on $C[t]$ and, in the consequence,
a solution of the positivity condition,
\begin{equation}\label{dmu}
a_n\ =\ \int_\R d\mu\ t^n.
\end{equation}
The state in Example also corresponds to a measure, namely to the
Dirac delta supported at $0\in \R$.
\medskip

On the other hand, combining the necessary homeomorphism
invariance condition (\ref{ball}) with the commutation relations
(\ref{[h,pi]}) allows to determine the expectation values for
products involving the quantum polymer position variables, and in
particular,  for arbitrary product $ \prod_{i=1}^n\,
\hat{h}_{k_i,x_i}\,\prod_{j=1}^m\hat{\pi}_{U_j}$ of the quantum polymer
position and momentum variables (including the degenerate case
$\prod_{i=1}^n\, \hat{h}_{k_i,x_i}$ ),
\begin{align}\label{prodh}
\langle \prod_{i=1}^n\, \hat{h}_{k_i,x_i}\,\prod_{j=1}^m\hat{\pi}_{U_j}
\,\rangle\ &=\ 0,
\ \ \ \ {\rm unless}\\
[\,\prod_{i=1}^n \hat{h}_{k_i,x_i},\,\hat{\pi}_U\,]&=0\ \ \ \
{\rm for\ \
 every}\ \ U\in\U.
\end{align}
The second case above takes place for example, when the product
involves the quantum polymer momentum variables only, or if
\begin{equation}
x_1=\ldots =x_n,\ \ \ {\rm and} \ \ \ \sum_{i=1}^nk_i\ =\ 0.
\end{equation}

In conclusion, we have formulated above the sequence
of  conditions necessary for a state $\langle\cdot\rangle$
defined on the polymer *-algebra $\kwa$ to be homeomorphism invariant:
{\it (\ref{prodpi}),  the existence of a state
$\mu$ on the polynomial algebra
such that (\ref{a_n}), and finally (\ref{prodh}).
If we additionally  assume that the state $\langle\cdot\rangle$
satisfies the condition (\ref{hh}),
then, the state is determined by the state
$\mu$ induced  on the
polynomial algebra $C[t]$.} Notice, however, that the {\it existence}
of the considered state $\langle\cdot\rangle$ has been being
assumed.

\section{The existence of the homeomorphism invariant states}
\label{existence}
 The existence statement we formulate now
has a conditional character. The unconditional  existence
statement we make next, will apply after imposing the {\it
semi-analyticity} condition on the subsets $U\subset M$ used to
label the momenta. Let us begin with the conditional statement:

{\it There exists a homeomorphism invariant state
$\langle\cdot\rangle$  on the polymer *-algebra $\kwa$
different the the state introduced in Example,
if and only if it is true that for every
finite sequence of balls\footnote{We call
$B\subset M$ a ball if there is a local chard
in the atlas of $M$ and a homeomorphism
$\varphi:M\rightarrow M$ such that
$\varphi(B)$ is mapped by the chart into a ball
in $\R^N$.} $B_1,\,\ldots,\,B_n\in M$
\begin{equation}\label{conjucture}
\sum_{i=1}^n m_i\one_{B_i}\ =\ 0\ \ \ \ \Rightarrow \ \
\ \ \sum_{i=1}^n m_i\ =\ 0,
\end{equation}
where given a subset $U\subset M$, $\one_U$ stands for the
characteristic function, and the coefficients $m_1,\,\ldots,\,m_n$
are integers. Moreover, if the state above exists, then the
conditions (\ref{prodpi}, \ref{a_n}, \ref{prodh},\ref{hh})
uniquely define a state $\langle\cdot\rangle$ for every given
state $\mu$.}
\medskip

The family  ${\cal U}$ of the subsets $U\im M$ labelling the
momenta has been   defined at the beginning of this section. If we
strengthen one of the defining conditions, and use instead the
family ${\cal U}^{\rm (sa)}$ of all the  open, bounded and  {\it
semi-analytic} subsets (assume for the simplicity $M=\R^N$)
\cite{loj} then all the considerations presented above continue to
be true   (The only add we have to make is defining the
homeomorphism invariance appropriately - see below). Moreover, it
is easy to prove that for the semi-analytic balls the implication
(\ref{conjucture}) is true! Therefore, the conclusions of
existence statement made above also becomes true unconditionally.

Using the semi-analytic sets breaks the usual diffeomorphism
symmetry. However, the  group of the so called \cite{lost} semi-analytic
diffeomorphisms which preserve $\U^{(sa)}$ is considerably
larger then the group of the analytic diffeomorphisms.
In particular, the group has local degrees
of freedom, in the sense that for every point $x\in M$
and its neighborhood $U'$, there is a semi-analytic
diffeomorphism of $\R^N$ which moves $x$, but restricted to
$\R^N\setminus U'$ is the identity map. It is not hard
to generalize the definition of the semi-analyticity to
a manifold $M$. It gives rise to exactly the
category of manifolds which are used  in LQG.

Denote by $\kwa^{\rm (sa)}$ the polymer *-algebra corresponding to
the labelling set ${\cal U}^{\rm (sa)}$. Of course $\kwa^{\rm
(sa)}\subset \kwa$.
 The homeomorphism invariance of a state on  $\kwa^{\rm (sa)}$
we need for our arguments to continue to be true
in the semi-analytic case is defined as follows.
 For every homeomorphisms
$\varphi:M\rightarrow M$,
\begin{equation}
\varphi*: \kwa^{\rm (sa)}\rightarrow  \kwa,
\end{equation}
and there may be a non-empty intersection $\varphi( \kwa^{\rm (sa)})\cap
\kwa^{\rm (sa)}$. Then, a state $\langle\cdot\rangle$ defined on the polymer
*-algebra $\kwa^{\rm (sa)}$ is called homeomorphism invariant if
\begin{equation}\label{homeo}
\langle\varphi(A)\rangle = \langle A\rangle,
\end{equation}
for every homeomorphism $\varphi$ and every
$A\in\varphi(\kwa^{\rm (sa)})\cap \kwa^{\rm (sa)}$.

\section{The corresponding GNS representations}\label{representations}
The states constructed in Sections \ref{necessary} and \ref{existence}
are finally used to find  representations of the
polymer *-algebra $\kwa^{\rm (sa)}$.
We will skip here the  GNS construction and
directly introduce the resulting representations.

Consider the vector space $\Cyl\otimes C[t]$ (the algebraic tensor
product) where the second factor is the space of polynomials of
one real variable familiar from Section \ref{necessary} (see it
for the notation), and $\Cyl$ is the set of the finite formal
linear combinations of elements $|\ka\rangle$ labelled by all the
maps $\ka:M\rightarrow \R$ of a {\it finite} support,
\begin{equation}
\Cyl\ =\ \{\sum_{i=1}^n a_i |\ka_i\rangle\,:\, n\in\N,a_i\in\C\}.
\end{equation}
Behind the family states we introduced in the previous section
there is a single underlying representation $\rho$ of the
polymer *-algebra  $\kwa^{\rm (sa)}$ in
the vector space $\Cyl\otimes C[t]$ and a {\it family} of hermitian
positive definite  forms.
The representation is defined by the following action of the
generators:
\begin{align}\label{rho}
\rho(\hat{h}_{k',x})(|\ka\rangle\otimes p(t))\ &=\
|\ka+k'\one_{\{x\}}\rangle\otimes p(t)\\
\rho(\hat{\pi}_U)(|\ka\rangle\otimes p(t))\ &=\
\sum_{x\in U}\ka(x)|\ka\rangle\otimes p(t)\ +\ \chi_{\rm E}(U)|\ka\rangle
\otimes t p(t),
\end{align}
where $\one_{\{x\}}$ is the characteristic function of the
1-element set $\{x\}$, and  the sum in the second equation has only
finitely many non-zero
terms.\footnote{The fact, that the definition above extends to the
representation
of the polymer *-algebra $\kwa^{\rm (sa)}$ is not trivial, and relies on the
properties of the Euler characteristic and on the implication
(\ref{conjucture}).}

Now, we relate $\rho$ with the state $\langle\cdot\rangle_\mu$
on the algebra $\kwa^{\rm (sa)}$ defined by:
 a fixed state $\mu$
on the polynomial algebra $C[t]$, and the conditions 
(\ref{prodpi}, \ref{a_n}, \ref{prodh},\ref{hh}) (with 
the symbol $\langle\cdot\rangle$ replaced by  $\langle\cdot\rangle_\mu$). 
We introduce a positive definite hermitian form $(\cdot|\cdot)_\mu$ in
the vector space $\Cyl\otimes C[t]$, and quotient
the vector space
by the subspace of the `zero norm' states. The hermitian form is the tensor
product $(\cdot|\cdot)_{\Cyl}\otimes (\cdot|\cdot)_{C[t]}$
hermitian form.
In $\Cyl$, the form $(\cdot|\cdot)_{\Cyl}$ is  defined by
\begin{equation}
(|\ka_{\bf 1}\rangle\,\,|\,|\ka_{\bf 2}\rangle)_{\Cyl}\ =\
 \left\{\begin{array}{ll} 1, {\rm if}\ \ \ka_{\bf 1}\ &=\ \ka_{\bf 2}\\
0, {\rm if}\ \ \ka_{\bf 1}\ &\not=\ \ka_{\bf 2} .\\
\end{array}\right.
\end{equation}

The hermitian form  $(\cdot|\cdot)_{C[t]}$ in  the vector space $C[t]$, on the
other hand, is defined by a state $\mu$ on the *-algebra algebra $C[t]$,
namely
\begin{equation}
 (P(t)|Q(t))_{C[t]}\ =\ \mu(\overline{P(t)}Q(t)).
\end{equation}
The resulting unitary Hilbert space is the completion of
the unitary space
\begin{align}
\hilb_\mu\ &=\
\Cyl\otimes (C[t]/J_\mu)\ \ \ {\rm where}\\
J_\mu\ &=\  \{P(t)\,:\,\mu(\overline{P(t)}P(t)) = 0\},
\end{align}
and the representation $\rho$ induces the representation
$\rho_\mu$ in $\hilb_\mu$. The representation $\rho_\mu$ is
related with the state $\langle\cdot\rangle_\mu$ defined
by $\mu$ and the conditions (\ref{prodpi}, \ref{a_n}, \ref{prodh},\ref{hh})
in the  familiar way
\begin{equation}
 \langle A\rangle_\mu\ =\ \left(\,|\one_{\emptyset}\rangle\otimes 1_{C[t]}
\,\big|\, \rho(A)\, |\one_{\emptyset}\rangle\otimes 1_{C[t]}\,\right)_\mu,
\end{equation}
where $\one_{\emptyset}$ is the identically zero function on $M$
and $1_{C[t]}$ is the identically $1$ function on $\R$, and
moreover the vector used to define the state is a cyclic vector, i.e.
\begin{equation}
\rho_\mu(\kwa^{\rm (sa)})\left(|\one_{\emptyset}\rangle\otimes 1_{C[t]}\right)
\ =\ \hilb_\mu.
\end{equation}

In \cite{kl} we have also analyzed properties of the
representation $\rho_\mu$ of the polymer *-algebra $\kwa^{\rm
(sa)}$. As it has been already mentioned, for every quantum
polymer position variable $\hat{h}_{k,x}$, the operator
$\rho_\mu(\hat{h}_{k,x})$ is unitary in $\hilb_\mu$. If we fix
$x\in M$ and vary $k\in \R$ including $k=0$, the corresponding
operators $\rho_\mu(\hat{h}_{k,x})$ form a 1-dimensional unitary
group. Given a quantum polymer momentum variable $\hat{\pi}_U$
labelled by $U$ such that $\chi_{\rm E}(U)\not=0$, the operator
$\rho_\mu(\hat{\pi}_U)$ is essentially self adjoint provided the
state $\mu$ is defined by a measure on $\R$ (see
(\ref{a_n},\ref{dmu})) such that $C[t]$ is dense in L$^2(\R,\mu)$.
Let us consider that class of states. Given two measures $\mu_1$
and $\mu_2$, the corresponding representations are equivalent, if
and only if the measures are equivalent (meaning: the families of
the measure zero sets coincide). Finally, given a measure $\mu$ on
$\R$, the corresponding representation $\rho_\mu$ is irreducible
if and only if $\mu$ is the Dirac measure supported at an
arbitrary appoint $t_0\in \R$. In that case case, $\hilb_\mu =
\Cyl$, and
\begin{equation}
\rho(\hat{\pi}_U)|\ka\rangle\ =\
\sum_{x\in U}\ka(x)|\ka\rangle\ +\ t_0\chi_{\rm E}(U)|\ka\rangle.
\end{equation}

In particular, the previously known
representation corresponding to the state defined in Example of
Section \ref{necessary}
is defined by $\mu$ being the Dirac delta supported at $t_0=0$.

\noindent{\bf Acknowledgments} We have benefited from discussions
with  Abhay Ashtekar, John Baez, Witold Marciszewski, Tadeusz
Mostowski, Andrzej Oko{\l}\'ow, Hanno Sahlmann, Thomas Thiemann and
Andrzej Trautman. The work was partially supported by the 
Polish Ministry of Science grants 1 P03B 075 29 and 1 P03A 015 29.



\begin{thebibliography}{99}
\bibitem{l1} Smolin L 2005 The case for background independence
[hep-th/0507235]
\bibitem{kl} Kami\'nski, W Lewandowski J, Oko\l\'ow A 2005 
Background independent quantizations: the scalar field II
\bibitem{t1} Thiemann T 1998 QSD V : Quantum Gravity as the Natural
Regulator of Matter Quantum Field Theories
 {\it Class. Quant. Grav.}  {\bf 15} {1281-1314}  [gr-qc/9705019]\\
Thiemann T 1998
 Kinematical Hilbert Spaces for Fermionic and Higgs Quantum
 {\it Class. Quant. Grav.} {\bf 15} 1487-1512  [gr-qc/9705021]
\bibitem{als} Ashtekar A, Lewandowski J, Sahlmann H 2003
Polymer and Fock representations for a Scalar field
{\it Class. Quant.Grav.} {\bf 20} L11-1 [gr-qc/0211012]
\bibitem{lqg}  Ashtekar A, Lewandowski J 2004
Background independent quantum gravity: A status report
{\it Class. Quant. Grav.} {\bf 21}, R53 [gr-qc/0404018]\\
Smolin L 2004  An invitation to loop quantum gravity
[hep-th/0408048]\\
Rovelli C. (2004): \textit{Quantum Gravity},
Cambridge: Cambridge University Press, 2004\\
 Thiemann T 2001  {\sl Modern Canonical Quantum
General Relativity.} Cambridge: Cambridge University Press, in
press, [gr-qc/0110034]
\bibitem{al2} Ashtekar A, Lewandowski L 1994 Representation Theory
of Analytic Holonomy C* Algebras
{\it Knots and Quantum Gravity} ed  Baez J C  (Oxford U Press, Oxford)
[gr-qc/9311010]
\bibitem{lost0}
Sahlmann H 2002 Some comments on the representation theory of the
algebra  underlying loop quantum gravity [gr-qc/0207111]\\
Sahlmann H 2002  When do measures on the space of connections
support the  triad operators of loop quantum gravity? [gr-qc/0207112]\\
Oko\l\'ow A  and Lewandowski J 2003 Diffeomorphism covariant
representations of
  the holonomy- flux *-algebra  {\it Class. Quant. Grav.} {\bf 20},
  3543--3568 [gr-qc/0302059]\\
Sahlmann H and Thiemann T 2003  On the superselection theory of
the Weyl
  algebra for diffeomorphism invariant quantum gauge theories,
  [gr-qc/0302090]\\
Sahlmann H,  Thiemann T 2003 { Irreducibility of the
Ashtekar- Isham- Lewandowski representation} [gr-qc/0303074]\\
Oko\l\'ow A,  Lewandowski J 2004
Automorphism covariant representations of
  the holonomy-flux *-algebra, {\it Class.\ Quant.\ Grav.}  {\bf 22} 657
[gr-qc/0405119]\\
Fleischhack C 2004
Representations of the Weyl algebra in quantum geometry
[math-ph/0407006]
\bibitem{lost} Lewandowski J,  Oko{\l}\'ow A,  Sahlmann H, Thiemann T 2005
{ Uniques of Diffeomorphism Invariant State on Holonomy- Flux
Algebras} [gr-qc/0504147]
\bibitem{loj} \L ojasiewicz, S. (1964): Triangulation of
semi-analytic sets.\ \ \ Ann. Scuola. Norm. Sup. Pisa {\bf 18},
449-474\\
Bierstone, E. and Milman, P. D. (1988):
Semianalytic and Subanalytic sets.\ \ \ Publ. Maths. IHES {\bf
67}, 5-42
\end{thebibliography}
\end{document}